\documentclass[twocolumn]{aastex6}

\pdfoutput=1

\shorttitle{Stellar Stream Around NGC~5907}
\shortauthors{Laine et al.}

\begin{document}

\title{Metallicity and Age of the Stellar Stream Around the Disk Galaxy NGC~5907}

\author{Seppo Laine\altaffilmark{1}, Carl J. Grillmair, and Peter Capak}
\affil{Spitzer Science Center-Caltech, MS 314-6, Pasadena, CA 91125, USA}

\author{Richard G. Arendt}
\affil{CRESST/UMBC/NASA GSFC, Code 665, Greenbelt, MD 20771, USA}

\author{Aaron J. Romanowsky\altaffilmark{2}}
\affil{Department of Physics and Astronomy, San Jos\'{e} State University, One Washington Square, San Jose, CA 95192}

\author{David Mart\'{i}nez--Delgado}
\affil{Astronomisches Rechen-Institut, Zentrum f\"{u}r Astronomie der Universit\"{a}t Heidelberg, M\"{o}nchhofstr. 12-14, D-69120 Heidelberg, Germany}

\author{Matthew L. N. Ashby}
\affil{Harvard-Smithsonian Center for Astrophysics, 60 Garden St., Cambridge, MA 02138, USA}

\author{James E. Davies}
\affil{Space Telescope Science Institute, 3700 San Martin Drive, Baltimore, MD 21218, USA}

\author{Stephen R. Majewski}
\affil{Department of Astronomy, University of Virginia, P.O. Box 400325, Charlottesville, VA 22904-4325, USA}

\author{Jean P. Brodie}
\affil{University of California Observatories and Department of Astronomy and Astrophysics, University of California, 1156 High Street, Santa Cruz, CA 95064, USA}

\author{R. Jay GaBany}
\affil{Black Bird Observatory, 5660 Brionne Drive, San Jose, CA 95118, USA}

\and

\author{Jacob A. Arnold}
\affil{University of California Observatories and Department of Astronomy and Astrophysics, University of California, 1156 High Street, Santa Cruz, CA 95064, USA}

\altaffiltext{1}{seppo@ipac.caltech.edu}
\altaffiltext{2}{University of California Observatories and Department of Astronomy and Astrophysics, University of California, 1156 High Street, Santa Cruz, CA 95064, USA}

\begin{abstract}

Stellar streams have become central to studies of the interaction histories of nearby
galaxies. To characterize the most prominent parts of the stellar stream around the
well-known nearby ($d$ = 17 Mpc) edge-on disk galaxy NGC 5907, we have obtained and
analyzed new, deep $gri$ Subaru/Suprime-Cam and 3.6\,$\mu$m  {\it
Spitzer}/Infrared Array Camera (IRAC) observations. Combining the
near-infrared 3.6 $\mu$m data with visible-light images allows us to use a long
wavelength baseline to estimate the metallicity and age of the stellar population
along a $\sim$~60 kpc long segment of the stream. We have fitted the
stellar spectral energy distribution (SED) with a single-burst
stellar population synthesis model and we use it to distinguish between the proposed
satellite accretion and minor/major merger formation models of
the stellar stream around this galaxy. We conclude that a massive minor merger
(stellar mass ratio of at least 1:8) can best account for the 
metallicity of $-0.3$ inferred along the brightest parts of the stream.

\end{abstract}


\section{Introduction}
\label{s:intro}

Major mergers, mergers that take place between galaxies of comparable mass (say
stellar or total mass ratios of 1:3 -- 1:5 or larger, with the ratio being the mass of
the lower mass galaxy over the mass of the more massive galaxy), have attracted
attention in the past \citep*[e.g.,][]{arp66,schweizer82,hibbard96,bush08} because
they have features that are easily detected even in shallow imaging observations.
However, major mergers may be red herrings when studying the evolution of typical
galaxies, many of which may not have undergone a major merger event since $z$ = 1
\citep[e.g.,][]{xu12}. Much more common  \citep[e.g.,][]{rodri15} are ``satellite
accretion events'' \citep*[stellar or total mass  ratio of mostly $\lesssim$~1:50;
e.g.,][]{deason16} that are mergers of a dwarf galaxy with a massive parent galaxy.
Evidence for such cannibalism has been found around our own Milky Way galaxy,
including the Sagittarius Dwarf stream \citep[e.g.,][]{ibata01b,majewski03}, Monoceros
Ring \citep{newberg02}, Anticenter Stream \citep{grillmair06b}, Orphan Stream
\citep{grillmair06a,belokurov07}, and Styx Stream \citep{grillmair09}. Such streams
have also been seen around the Andromeda galaxy \citep[e.g.,][]{ibata01a}. 

The effects of mergers intermediate in strength between major mergers and satellite
accretion events, a.k.a. ``minor mergers'' (stellar or total mass ratios of roughly
between 1:50 and 1:5), have proven to be much harder to observe directly. Such mergers
are not expected  to transform galaxies as drastically as major mergers (disk to
elliptical transformation), but on the other hand one would expect the effects of such
mergers to show up as more  than just tenuous stellar population changes in the halo
(cf. Local Group streams mentioned above).
Moreover, \citet{stewart09} have argued that every large galaxy has undergone at least
one minor merger during its lifetime. Thus, it can be argued that it is the minor
mergers that are the most dynamically significant  merging events, in view of their
relatively high frequency (see also \citeauthor{zaritsky97} \citeyear{zaritsky97},
\citeauthor{bullock05} \citeyear{bullock05}) and the substantial dynamical and
structural impacts on the primary galaxy that such interactions can bring about. These
impacts include the heating and thickening of host galaxy disks, growth of galactic
bulges, hierarchical build-up of galaxy mass, counterrotating cores in host galaxies,
and triggering and maintenance of bar and spiral structures
\citep*[e.g.,][]{mori08,purcell10}. Observational evidence  for minor mergers has been
surfacing during the last decade, as fairly high surface brightness stellar streams
around several nearby galaxies have been detected in ground-based  visible-light
observations \citep[e.g.,][]{shang98,delgado08,delgado09,delgado10,duc15,jennings15}.

\begin{figure*}
\centering
%\epsscale{0.85}
\includegraphics[angle=270,scale=0.7]{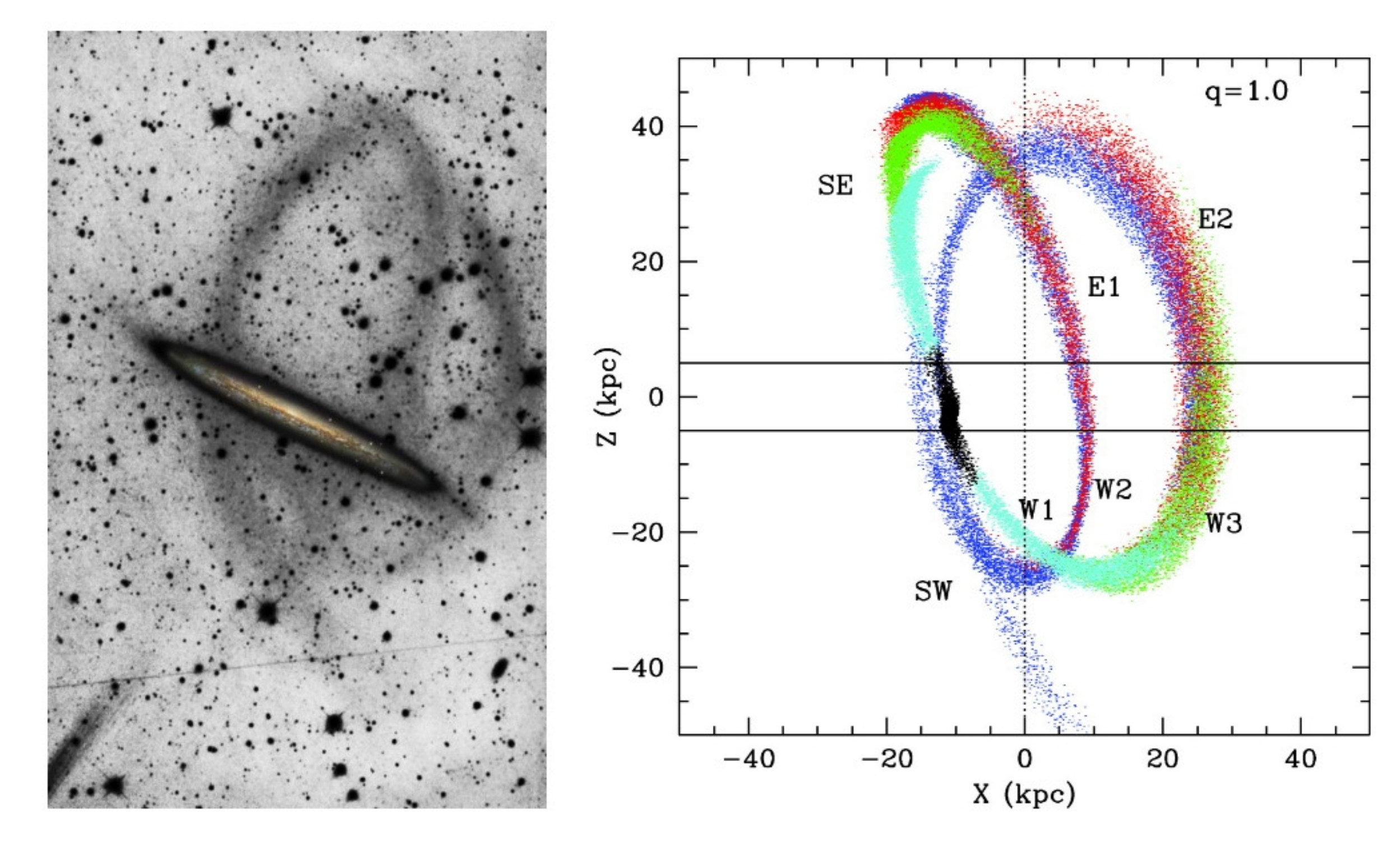}
\caption{Observed (left) and modeled (right) stellar stream around NGC 5907,
reproduced from Figures 2 and 4 of ``The Ghost of a Dwarf Galaxy: Fossils of the
Hierarchical Formation of the Nearby Spiral Galaxy NGC 5907'' by D.
Mart\'{i}nez--Delgado, J. Pe\~{n}arrubia, R. J. GaBany, I. Trujillo, S. R. Majewski,
\& M. Pohlen, ApJ, vol 689, issue 1, (2008) pp. 184--193. The image on the left is a
combined clear luminance (350--850 nm), red, blue, and green filter image, with a
color image made from the red, blue, and green filter images superimposed on the
saturated disk (see \citeauthor{delgado08} \citeyear{delgado08} for more information),
observed at the BlackBird Remote Observatory's 0.5-m telescope. The image on the right
is an N-body minor merger model from \citet{delgado08}. The different colors designate
particles that become unbound after different pericenter passages, with cyan, green,
red, and blue indicating particles that became unbound during the first, second,
third, and fourth most recent pericenter passages. The various letter designations are discussed in \citet{delgado08}. q = 1.0 refers to a spherical halo model. The image
on the left is about $17\farcm 5$ $\times$ $27\farcm 5$ in size. North is to the right
and east is up.\label{visfig}} 
\end{figure*}

NGC~5907 is a nearby ($d$ = 17 Mpc; \citeauthor{tully13} \citeyear{tully13}, AB$_{\rm
mag}$$^{3.6}$ = $-21.9$; \citeauthor{irsa}, $V_{\rm rot}^{\rm max}$ = 240
km~sec$^{-1}$; \citeauthor{casertano83} \citeyear{casertano83}) edge-on Sc-type disk
galaxy with a stellar stream around it, discovered by \citet{shang98}, and studied
further by \citet{zheng99} and \citet{delgado08}. The outer disk of NGC~5907 also has
a warp \citep[e.g.,][]{shang98}. \citet{delgado08} obtained very deep images of
NGC~5907 and revealed the spectacular long stellar stream (Figure~\ref{visfig}).
However, because they used a luminance filter \citep[see Fig. 1 in][]{delgado15},
\citet{delgado08} were not able to measure colors in the stream. \citet{delgado08}
also produced a simple N-body model that mimicked the observed loopy stellar stream as
a fossil of the tidal disruption of a single satellite in a merger event (total mass
ratio of 1:4000), rejecting the hypothesis of multiple merger  events in the halo of
NGC 5907. 

In contrast, \citet{wang12} reported a model that reproduced the general structure of
the stellar stream around NGC~5907 in a major merger scenario.  \citet{wang12}
suggested that the colors of the stream may be used to distinguish between a minor
satellite accretion event and a major merger origin of the stellar stream in NGC~5907 
(the stream is too faint to be observed spectroscopically). They used both 1) the
color-inferred iron abundance in the disk outskirts (affected also by stars in the
inner halo) and found that the $R-I$ color from \citet{zheng99} is similar to that of
massive disk galaxies instead of dwarf satellite galaxies, and 2) the
mass--metallicity relation in which, for example, the low-mass
Sagittarius satellite galaxy (with low [Fe/H]) would produce a  bluer color than what
is observed. We exploit a similar technique in this paper. We have obtained new deep
Subaru/Suprime-Cam $gri$ and {\it Spitzer}/Infrared Array Camera
(IRAC) 3.6\,$\mu$m images of NGC 5907 to measure the spectral energy distribution
(SED) and color indices of the brightest parts of the stellar stream east of the disk
of NGC 5907, and to compare them to the predictions of satellite
galaxy accretion and minor/major merger models. To our knowledge this is the first
time that IRAC data have been used to constrain the metallicity and the age of the
stellar population in a stellar stream around a  nearby galaxy (outside the Local
Group).

This paper is organized as follows. In Section~\ref{obs} we summarize the new
observations and data, in Section~\ref{magcolor} we present the surface brightness and
color index results and study the potential effects of an extended
point spread function (PSF), while Section~\ref{sed} addresses how we fit
single burst SED models from the flexible stellar population
synthesis (FSPS) code. We discuss our results in Section~\ref{discus} and present our
conclusions in Section~\ref{concl}.

\section{Observations and Data Reduction}
\label{obs}

\subsection{{\it Spitzer}/IRAC Observations and Data Reduction}
\label{iracred}

\begin{figure*}
\centering
%\epsscale{0.85}
\includegraphics[angle=0,scale=0.8]{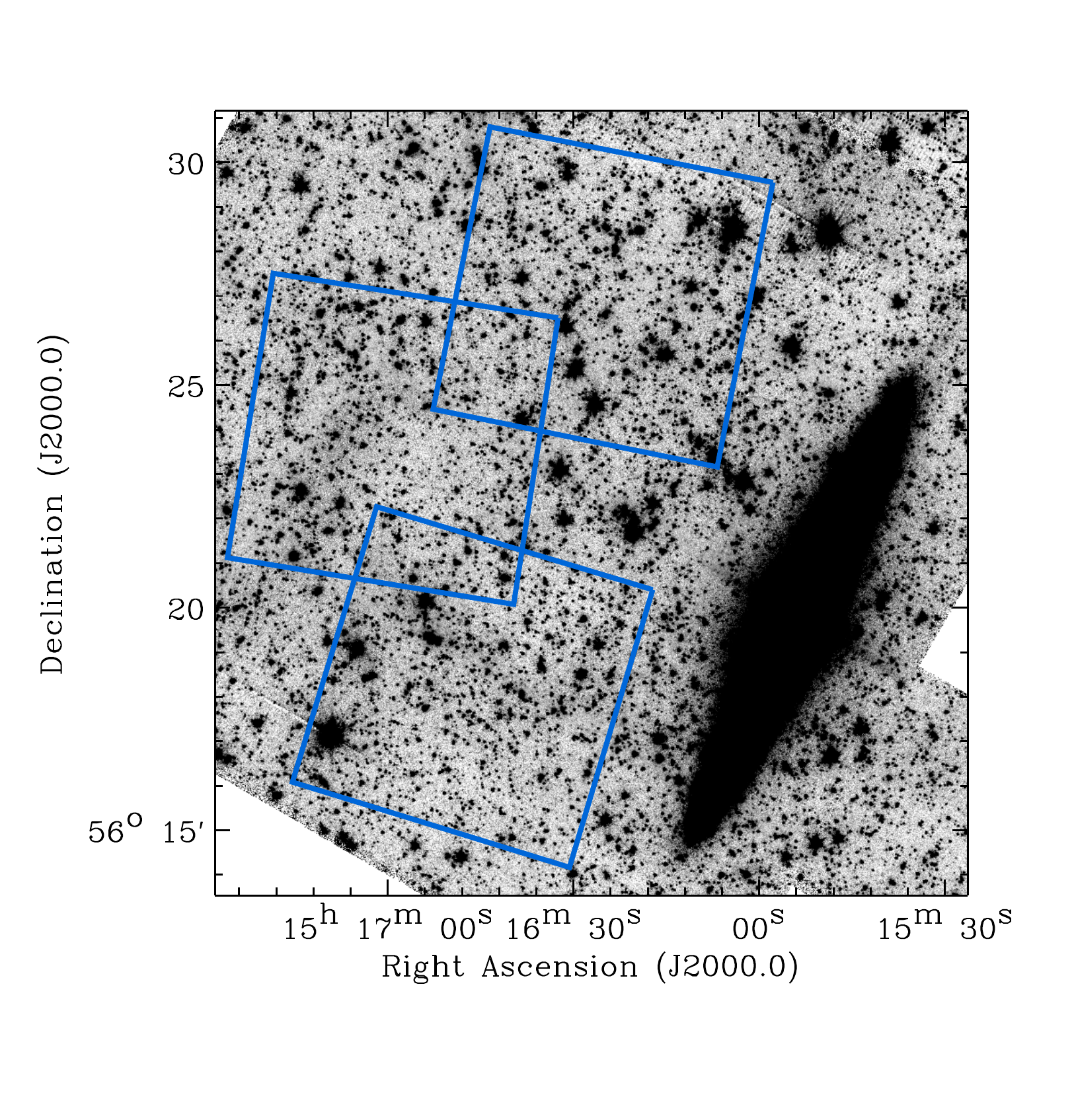}
\caption{Overlay of the mean location of all the IRAC frames in our new observations
on top of an IRAC 3.6 micron image from earlier NGC 5907 observations in {\it Spitzer} program 3. The boxes drawn have been expanded to show the full area covered by all dithered frames of the three new deeper pointings.\label{overlayfig}} 
\end{figure*}

{\it Spitzer}/IRAC \citep{werner04,fazio04} observed NGC 5907 at 3.6 $\mu$m on 2010
July 12, 14, and 20. We used three separate pointings of IRAC's 5\farcm 2 $\times$
5\farcm 2 field of view (observation identification numbers or
AORKEYs 34781952, 34782208, and 34782464 in program 60088; see
Figure~\ref{overlayfig}) that covered the brightest parts of the stream to the east
of the galaxy disk. We used 30 positions in a custom offset Reuleaux triangle pattern
\citep*{arendt00} with a scale of about 75 arcseconds and performed a five-point,
small-scale (25 arcseconds) cycling dither pattern around each point in the triangle
for each of the three pointings. The frame time used was 100 seconds (which gives 93.6
seconds of exposure time per frame at 3.6 $\mu$m). We consequently spent 14,040
seconds per field on the stream. Previous data from {\it Spitzer}
program 3 with ten 100 second frames had hinted at the brightest
parts of the stellar stream (see Figure~\ref{overlayfig}). Our
new observations are a factor of 15 longer in exposure time.

Production of the final mosaiced images made use of several tools. The processing
started with the corrected basic calibrated data (CBCD) frames
produced by version S18.18.0 of the IRAC pipeline. The data for each astronomical
observing request (AOR; corresponding to a single dithered field-of-view) were then
self-calibrated using the procedures developed by \citet{fixen00} to derive
delta-corrections to the detector offsets and to fit time-dependent drifts in the sky
intensity (whether due to instrumental effects or changes in the zodiacal light). We
then ran the data through a custom column pulldown artifact corrector written by one
of us (M. Ashby). The data were again self-calibrated using all three AORs together to
derive delta-corrections to the offsets (essentially zero at this point) and the best
fit to time-variable offsets. These variable offsets optimize the background matching
where there is partial overlap between the three fields.

Subsequently, to obtain the best astrometric solution, we ran the data through one
iteration of the Great Observatories Origins Deep Survey (GOODS) pipeline
\citep{grumm05} artifact mitigation. Astrometry in the processed CBCD frames was
refined using the standard {\it Spitzer} Science Center (SSC) provided distortion
solution. Rotation and translation parameters for all individual CBCD frames were
calculated after distortion had been removed. The rotation and translations were based
on both internal references (same source in multiple CBCD frames) and external
references in the form of Sloan Digital Sky Survey sources covering the whole NGC 5907
field. More weight was given to the internal match so that our PSFs are very
sharp. The result is that the astrometry in our frames is tied to Sloan Digital Sky
Survey (SDSS) astrometry.

We then used the GOODS pipeline mosaicer to mosaic the modified CBCDs. Remaining
cosmic ray hits were removed using outlier rejection when the CBCD frames were
combined into a mosaic. Some differences of the GOODS pipeline mosaicing compared to
the standard SSC pipeline mosaicer include 

\begin{enumerate}

\item less astrometric uncertainty (because of the very sharp PSFs) in the mosaics
coming from the GOODS pipeline. We weighted internal matches (particular sources
seen on many CBCD frames) as indicated above. This was important when we removed
sources by PSF fitting below; 

\item the drizzle algorithm used to reconstruct an image from undersampled pointings 
gives superior resolution compared to {\it Spitzer} custom software mosaicer and
point source extractor (MOPEX) mosaics built with the default MOPEX mosaicing method.

\end{enumerate}

As stars at the distance of NGC 5907 are completely unresolved in our images, we 
removed all detected point sources. We ran the SSC-provided custom software
astronomical point source extractor for MOPEX (APEX) to find point sources and
removed them using the APEX quality assurance (QA) pipeline. We opted to remove
sources that were centrally concentrated (aperture flux density ratio within a ten
pixel to a three pixel radius was less than 1.3) or very faint sources that turned out
to be essentially point-like. Once we had removed the point sources via PSF fitting,
we still had the extended sources left. We used SExtractor \citep{bertin96} to mask
the extended sources. We extended the masks by two pixels in radius in Adobe
Photoshop to remove a visible halo around the extended sources. Because the IRAC 3.6
$\mu$m PSF is fairly steep within the core (it drops by a factor of about 500 within a
5-arcsecond radius), the outlying parts of the PSF are well within the noise in the
mosaic image for all but the brightest sources (see also Section~\ref{psf} below).
Typical slightly extended background sources in the IRAC images have a central pixel
value of about 0.05 MJy/sr while the 1-$\sigma$ noise in the final mosaic is about
9$\times$10$^{-4}$ MJy/sr, and the brightness of these sources at 3.6 $\mu$m drops
below the noise level within about two native pixels or about $2\farcs 5$. The
amount of light from extended PSFs due to the bright edge-on galaxy disk and the
brightest stars in the field of view is further discussed in Section~\ref{psf} below.

\subsection{Subaru Observations and Data Reduction}
\label{subarured}

\begin{figure*}
\centering
%\epsscale{0.85}
\includegraphics[scale=0.7]{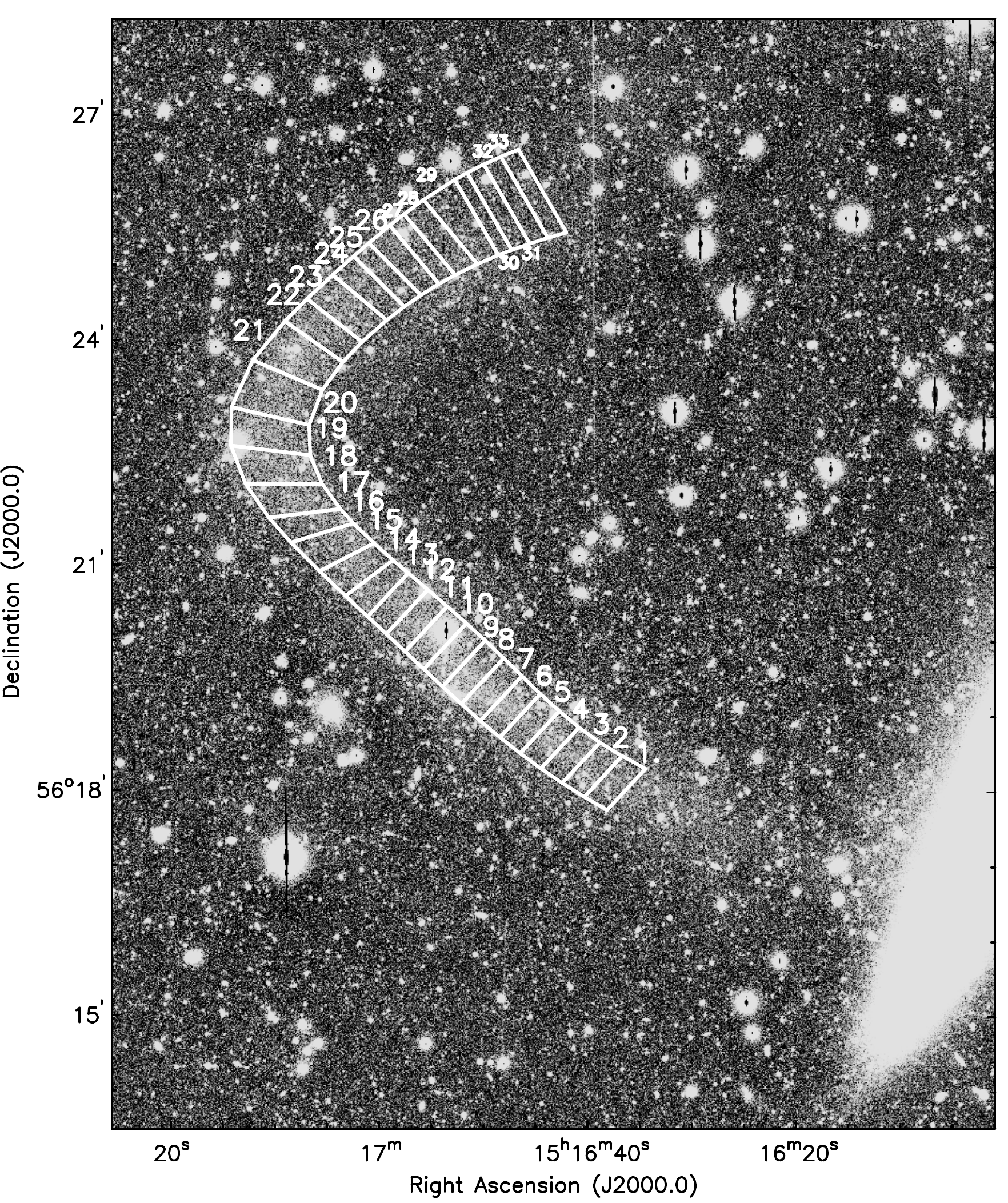}
\caption{Photometry apertures on top of a Subaru Suprime-Cam $r$-band image. The 
apertures in which the surface brightnesses and colors were measured are shown on 
top of an $r$-band Subaru Suprime-Cam image of the stellar stream in NGC 5907. The
apertures correspond to roughly 4--5 kpc $\times$ 2--3 kpc each. Even though in this 
$r$-band image the stream continues to the southwest (SW) of aperture 1, the surface brightness was so low there in the other bands that no attempt was made to measure the stream surface brightness SW of aperture 1.\label{boxfig}} 
\end{figure*}

\begin{figure*}
\centering
%\epsscale{0.85}
\includegraphics[angle=270,scale=0.6]{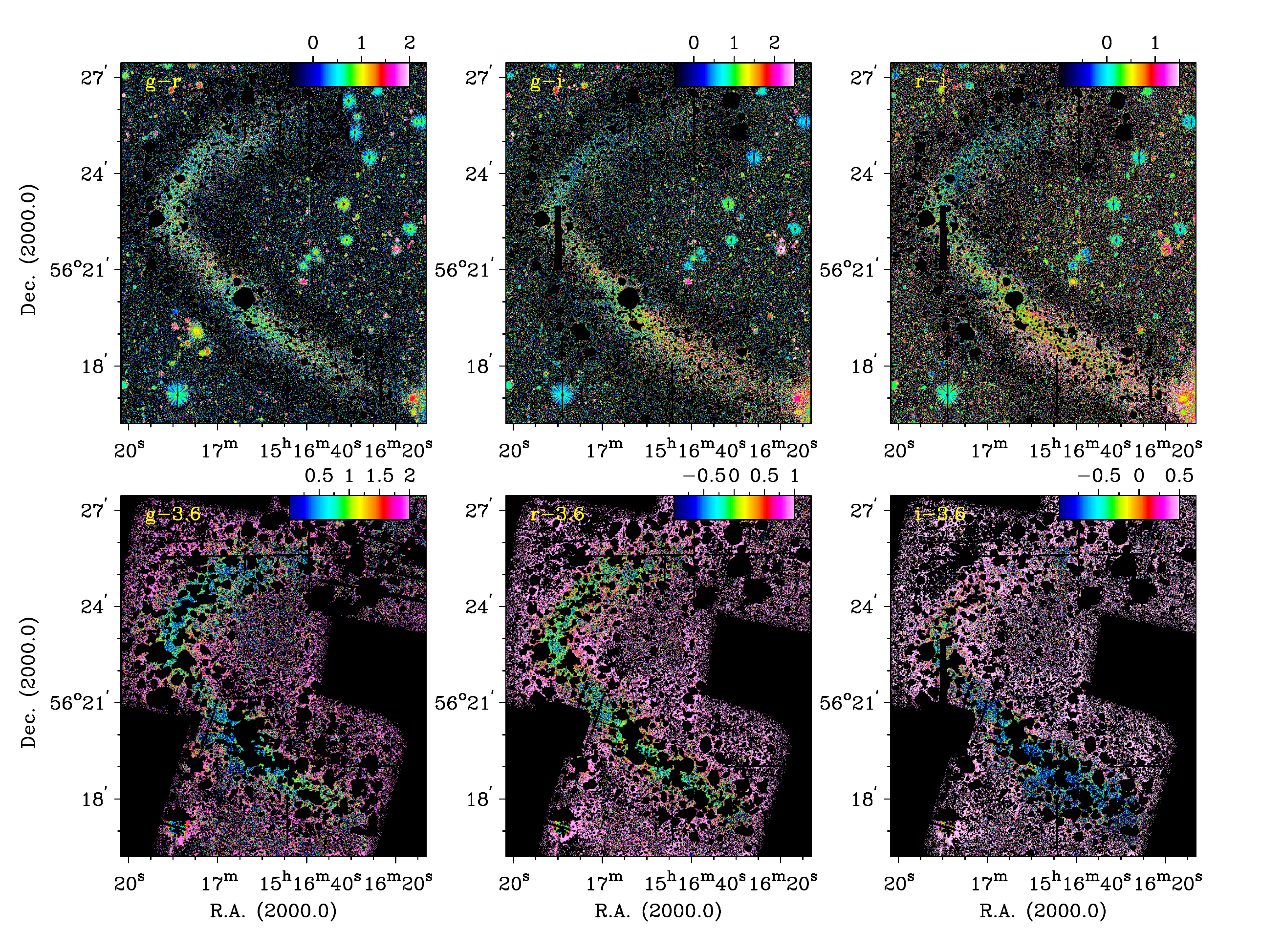}
\caption{NGC 5907 color index images. The color index images $g-r$, $g-i$, $r-i$,
$g-3.6$, $r-3.6$, and $i-3.6$ (in AB magnitudes) for the NGC 5907 stellar stream
after  regridding to 1 arcsecond $\times$ 1 arcsecond pixel size are shown. The units
on the color scales are the color magnitudes (per pixel), as shown by the color bar at
the top of the images. Color gradients can be seen in the visual color index images
(top row) most clearly, the color changing from reddish to bluish from bottom to top
along the stream. See the  discussion of these gradients in
Section~\ref{estmagcolor}.\label{colorindexfig}} 
\end{figure*}

\begin{figure*}
\centering
%\epsscale{0.85}
\includegraphics[angle=270,scale=0.6]{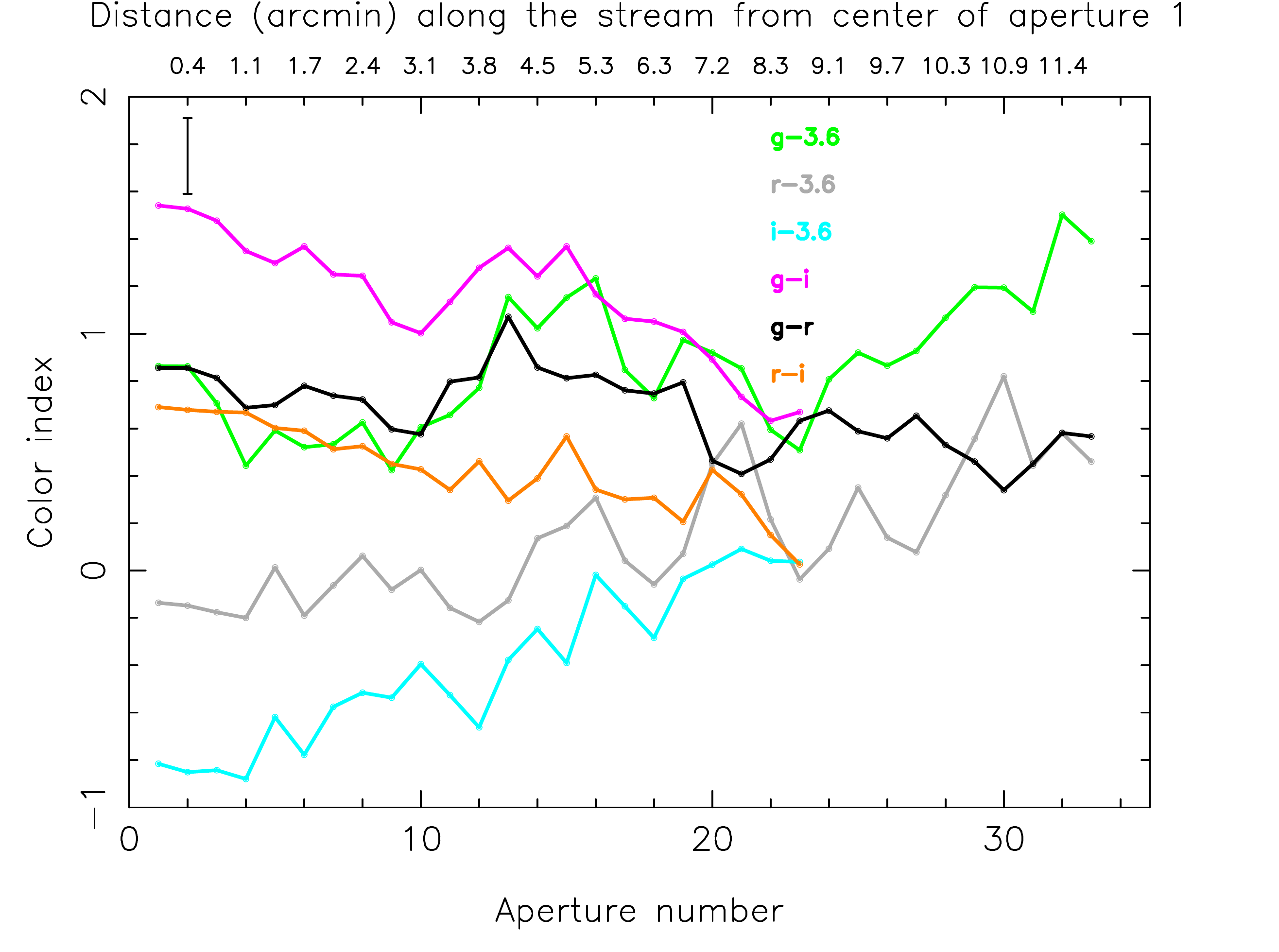}
\caption{Colors along the stellar stream in NGC 5907. This image shows the AB magnitude colors measured along the stellar stream around NGC 5907 in the apertures shown in Fig.~\ref{boxfig}. The lowest aperture numbers are closest to the galaxy disk, as can be seen in Figure~\ref{boxfig}. A representative total error bar is drawn in the upper left corner. The uncertainties are much larger for apertures beyond aperture 23 where the stream becomes much fainter. The statistical uncertainties were estimated by dividing the rms in the individual apertures by the square root of the number of independent beams in them.\label{magcolorfig}} 
\end{figure*}

\begin{figure}[b]
\includegraphics[width=3.5in]{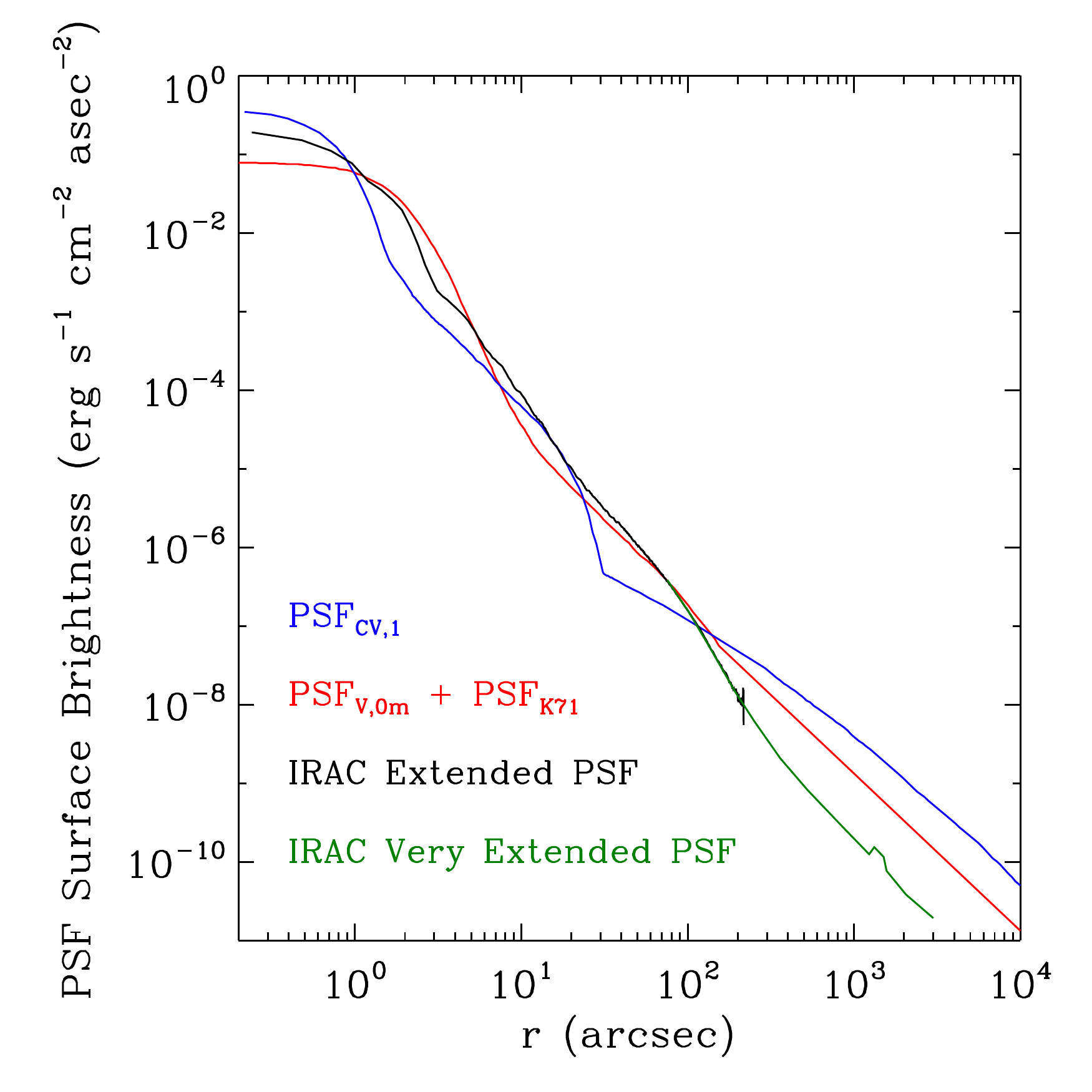}
\caption{Radial profiles of all the PSFs that were used to model the extended PSF background of NGC 5907. The black and green lines show the IRAC PSF derived from the warm extended PSF
and from $\beta$ Gru observations respectively. The red and blue profiles are the two 
adopted profiles \citep{sandin14} used with the Subaru data.\label{fig:psf}}
\end{figure}

\begin{figure}[h]
\includegraphics[width=3.5in]{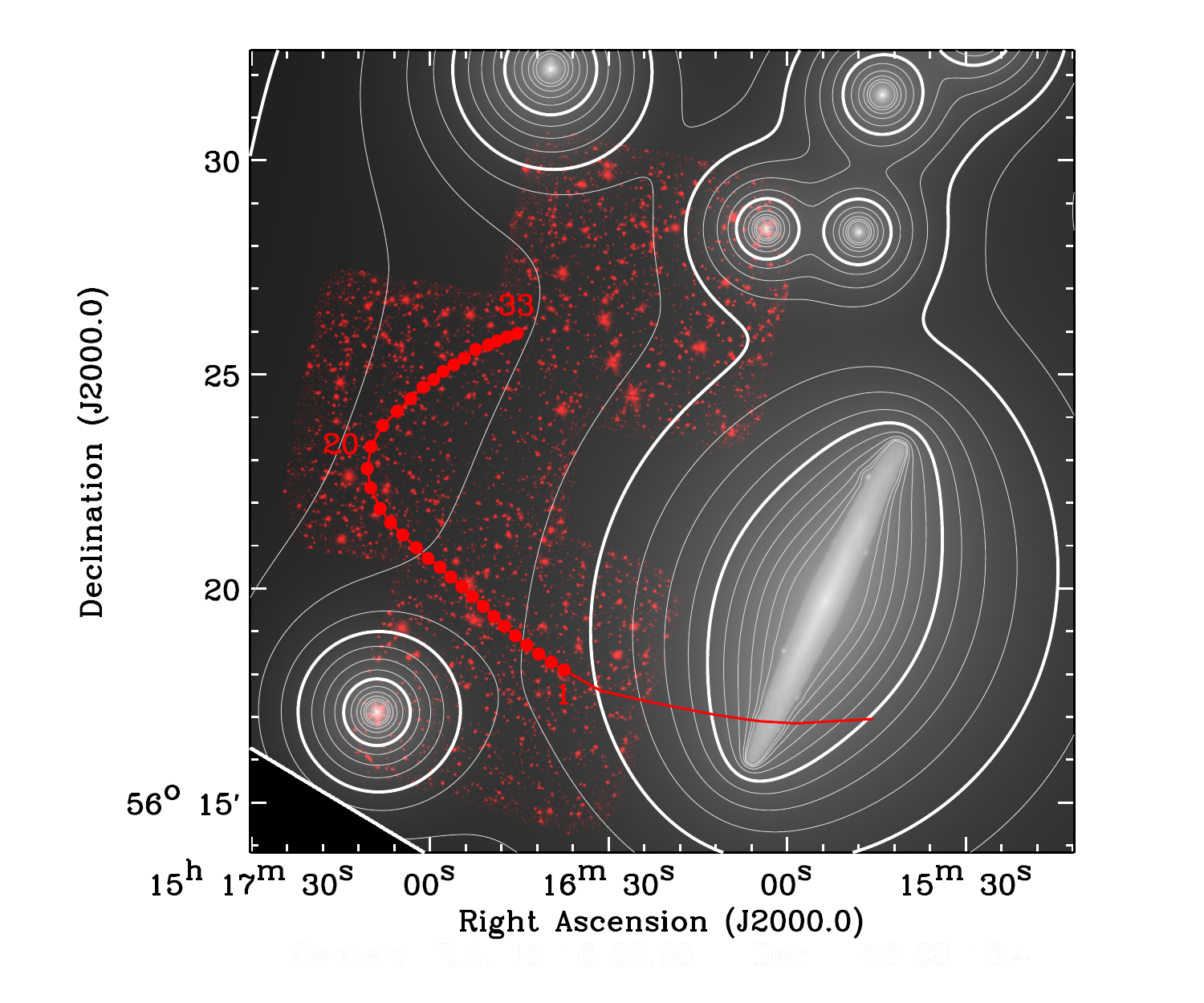}
\caption{IRAC 3.6 $\mu$m extended PSF background model. Contours
are logarithmically spaced with thick contours at $10^{-5}$, $10^{-4}$, and $10^{-3}$
MJy~sr$^{-1}$. The surface brightness of the stream is at least a factor of ten 
higher than that of the extended PSF background light. The centers of the apertures along the stream are indicated by bright red dots. The red line indicates the longer slice depicted in Figure \ref{fig:slice}. The deep IRAC image is shown in red.\label{fig:model1}}
\end{figure}

\begin{figure*}[h]
\includegraphics[width=3.5in]{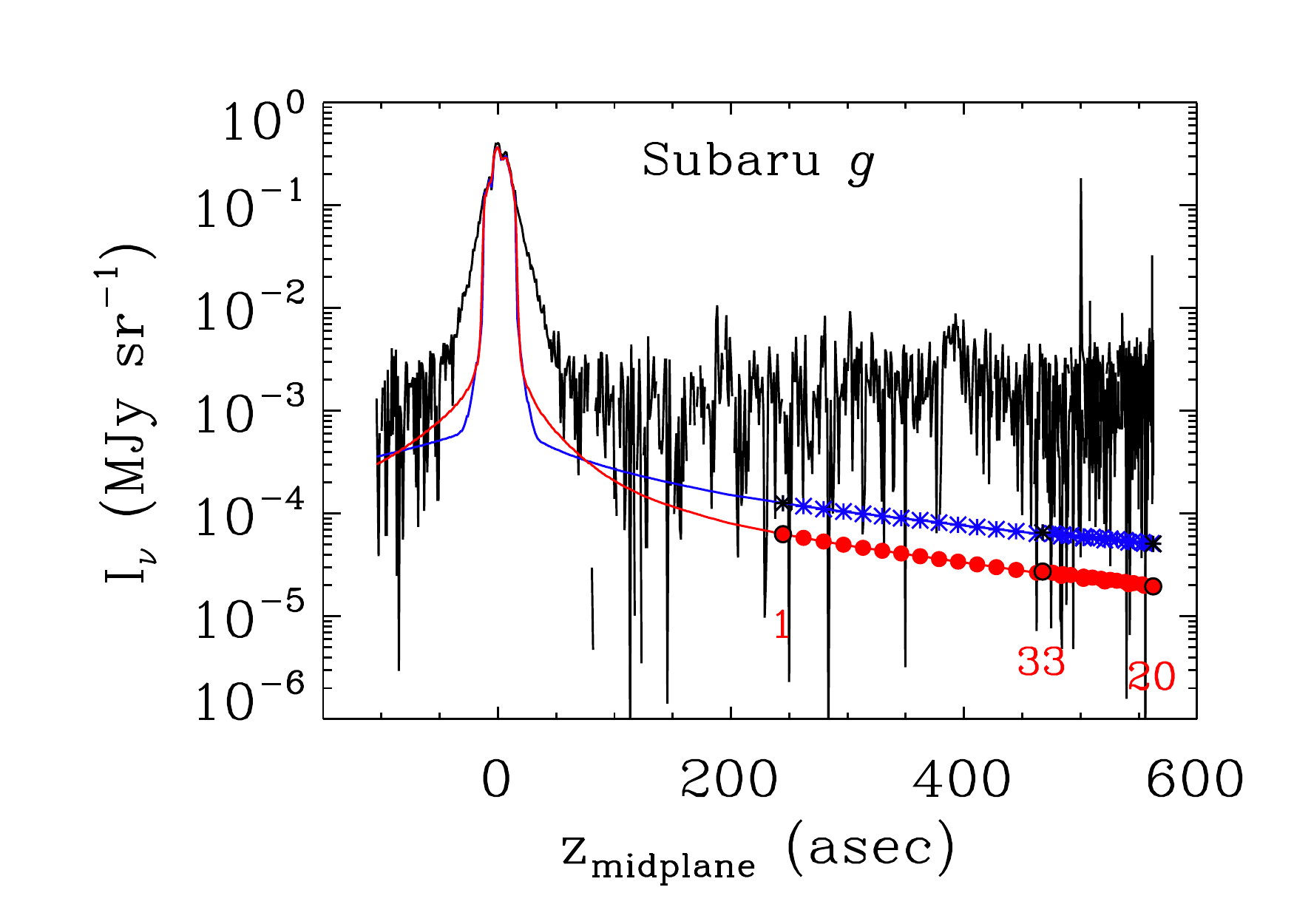}
\includegraphics[width=3.5in]{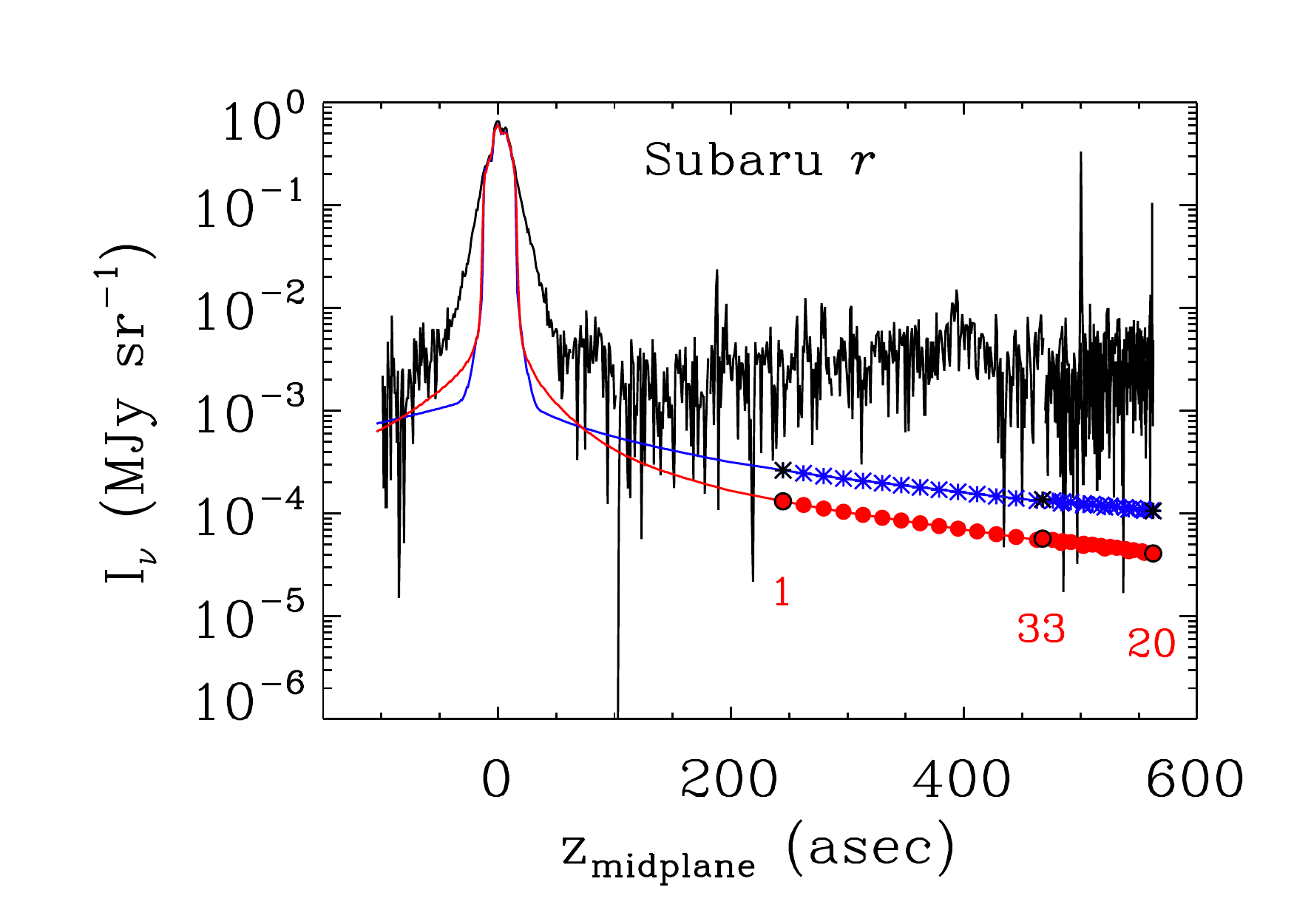}\\
\includegraphics[width=3.5in]{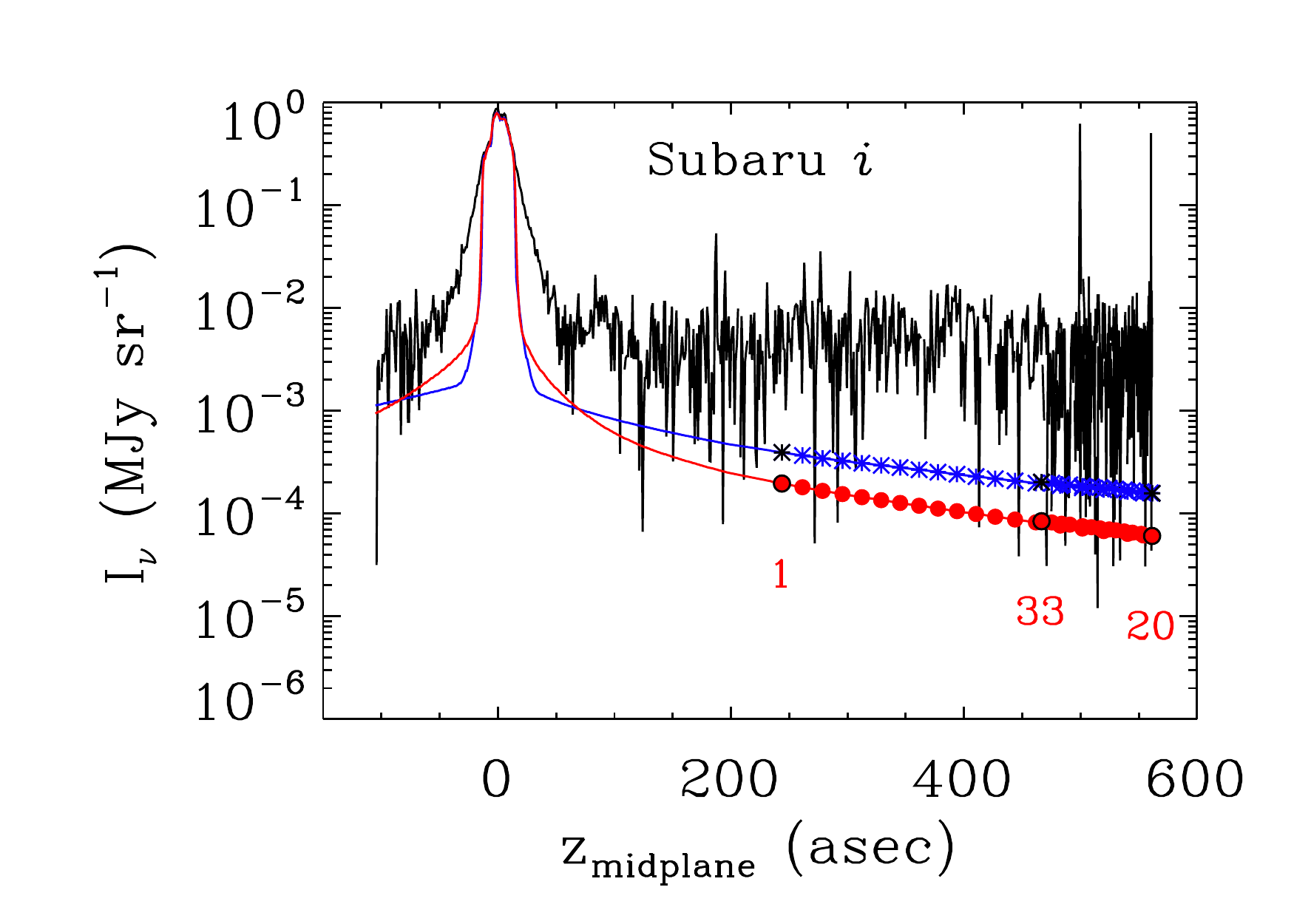}
\includegraphics[width=3.5in]{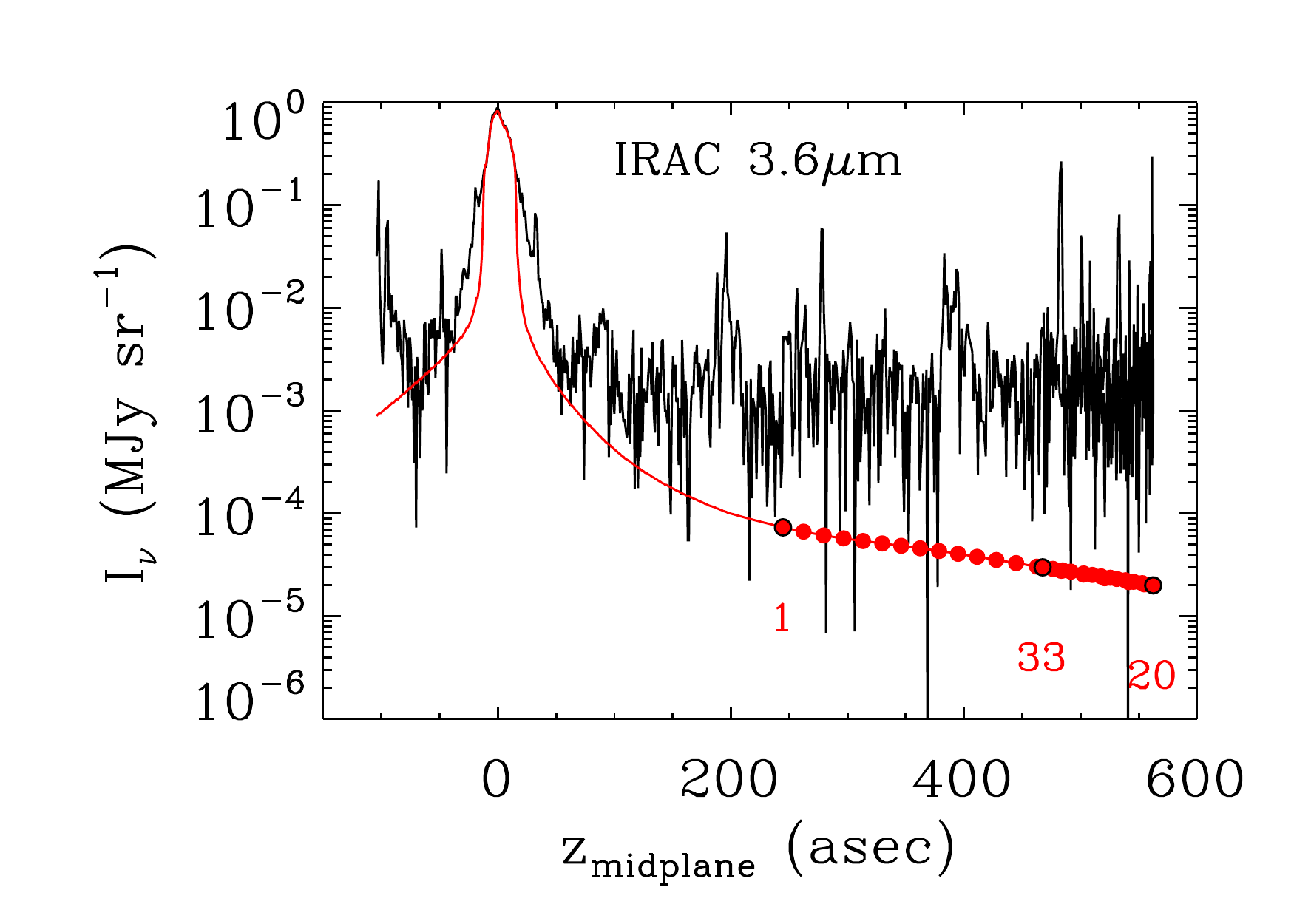}
\caption{Slices (as shown in Fig.~\ref{fig:model1}) through the data (black) and
extended PSF background models (blue and red colors) crossing  the plane of NGC 5907
and along the stellar stream. The slices are plotted as a function of distance from
the galaxy midplane. For left to right and top to bottom are Subaru $g$, $r$, and
$i$, and IRAC 3.6 $\mu$m bands. For the Subaru model slices, the 
PSF$_{\rm{V,0m}}$+PSF$_{\rm{K71}}$ model is red and the PSF$_{\rm{CV,1}}$ model is
blue. The bumps near region 10 are from a star that was masked out before the
measurements were made. Red numerals indicate aperture identifiers as shown in
Fig.~\ref{boxfig}.\label{fig:slice}}
\end{figure*}

NGC 5907 was observed using the Suprime-Cam imager \citep{miyazaki02} consisting of
ten 2048 $\times$ 4096 pixel charge-coupled devices (CCDs) with a total field of
view of 34\arcmin~$\times$~27\arcmin\ covered by 0\farcs 20 pixels, mounted at the
prime focus of the 8.2 m Subaru telescope, on the morning of 2010 April 12. We
obtained images  through the SDSS $gri$ filters in photometric conditions with 0\farcs
6 -- 0\farcs 9 seeing. Between four and five exposures were obtained per filter to
help with cosmic ray rejection and to increase the dynamic range of the final co-added
mosaics. Additionally, the telescope pointing was dithered slightly between exposures
to account for bad pixels/columns and to fill the gaps between individual CCDs.

The fully depleted charge-coupled device (FDCCD) detectors installed on Suprime-Cam in
January 2009 are each divided into four regions read out by separate sets of
electronics with different bias levels and gains. We did not use the standard pipeline
reduction for the Subaru Suprime-Cam data because it is important for us to remove all
the sources of underlying background emission (whether  instrumental or from the sky)
as carefully as possible. In our reduction the bias was first subtracted using the
overscan area of the images as defined in the headers and the gain was scaled to 1.0
(everything converted to electrons) for all regions on all chips. Then, a median bias
image was created from a set of bias frames and subtracted to remove residual
structure from the edges of the readout regions. Next, a median dome flat was
constructed for each band, and divided out of the data frames.  This left images that
were largely flat in the central region of the camera, but with significant large
scale structure in the sky due to scattered light in the frames, and to a lesser
degree in the dome flats \citep{capak07}.

To remove the sky background, a median sky image was created from the data frames in each band. To avoid over-subtracting extended structure, a two-step masking process was used. First, object detection was run on each frame and used to make a preliminary mask that, in turn, was used to make a first pass sky image. This was then scaled to the background of each exposure and subtracted from the images. This first pass sky-subtracted image was then combined into a mosaic, and a manual mask was made of all extended emission regions visible in the mosaic, ensuring that we are not accidentally removing parts of the stream. The manual mask was then propagated back to the individual images, and a second pass sky frame made. This second pass sky frame was then inspected, noisy regions masked, and fitted with a tessellated surface. The tessellated surface was then scaled to the background in each exposure and subtracted.

Finally, to remove scattered light that varies from exposure to exposure, each sky-subtracted exposure was masked with both the detected object mask and the extended source mask, divided into a 8$\times$16 grid, and fitted by a third-order two-dimensional polynomial over that grid, which was then subtracted. After attempting fits with higher and lower order polynomials, we settled on the third order polynomial. That fit to the grid successfully removed the scattered light (that varies only on large scales, $\sim$~7$\arcmin$$\times$14$\arcmin$) in the center of the frames, but did not subtract the extended emission. After this process some residual scattered light was visible around the edge of the frame where the scattered light had more structure \citep{capak07}. 

We used the following color-corrected magnitude zero points to convert the counts into
magnitudes: $g_{zp}$ = 35.95, $r_{zp}$ = 36.32, and $i_{zp}$ = 36.19 for individual
frame exposure times of 203, 119, and 142 seconds in $g$, $r$, and $i$, respectively.
These zero-points are good to 0.002 mag random error for $g$ and $r$ and to 0.003 mag
random error for $i$ relative to SDSS, using the color conversions in \citet{capak07}
that are valid for colors $-1$~$<$~$g$$-$$r$~$<$~1 for $g$ and $r$ and
$-1$~$<$~$r$$-$$i$~$<$~1 for $i$. The zero point calibration was done against SDSS
Data Release 8 with SExtractor source extractor auto magnitudes, using 453, 455,
and 525 stars in $g$, $r$, and $i$, respectively.

We masked out extended objects by hand in the stream in the $g$, $r$, and $i$
images. We then cross-convolved the Subaru images with the IRAC PSF and the IRAC
3.6  $\mu$m image with the Subaru PSFs. The Suprime-Cam $gri$ PSFs were generated by
SSC's pfr\_estimate.nl Perl script. We subsequently applied a kernel to taper the
PSFs to avoid ringing in the cross-convolved images. Color index images were then made by
combining the mosaiced images from the various observed bands.

\section{Surface Brightnesses and Color Indices}
\label{magcolor}

\subsection{Estimation of Colors and Surface Brightnesses}
\label{estmagcolor}

We analyzed the color index images by constructing apertures, shown in
Figure~\ref{boxfig}, of the stream, and measured the median color of the color index
images (Figure~\ref{colorindexfig}). The resulting color plots 
(Figure~\ref{magcolorfig}) show the median colors vs. the number of the aperture as 
shown in Figure~\ref{boxfig}. Mean colors and several other properties of the stream
are also shown in Table~\ref{table1}. We tested the robustness of the colors by making
Subaru Suprime-Cam images that used the first pass sky-subtracted image (see
Section~\ref{subarured}) in the background subtraction (colors changed by less than
0.05 mag) and by increasing the masked areas within the stream by 5--10 pixels on each
side (colors changed by less than 0.1 mag). Finally, by measuring the variance in
``empty'' sky regions outside the stream we estimated the uncertainty in the sky
background value subtraction in the IRAC image to be about 0.08 mag. Therefore, we
conservatively believe that our measured colors are accurate to 0.2 mag. Extinction
corrections due to the Milky Way are expected to be less than 0.02 mag in the visual
colors and even less for the visual -- infrared colors, and therefore have not been
applied, as their effect is much smaller than the other uncertainties in the colors.

In general the surface brightness is highest in apertures 5 -- 10 and decreases in
both directions outside those apertures. Specifically, the surface brightnesses peak
at about 27.4, 26.7, 26.1, and 26.7 AB mag~arcsec$^{-2}$ in $g$, $r$, $i$, and
3.6 $\mu$m, respectively, and are at their faintest towards the  northern end of the
detected area at 28.5, 28.0, 27.8, and 27.6 AB mag~arcsec$^{-2}$ in $g$, $r$,
$i$, and 3.6 $\mu$m, respectively. The brightest parts of the stream have a
transformed $R-I$ color of about 0.7 using the transformation equations of
\citet{jester05} ($R-I$ $\approx$ $r-i$ + 0.2), which is in the range of earlier
estimates of $R-I$ = 0.5 \citep{zheng99} to 0.9 \citep{shang98}  in the brightest part
of the stream (although note that they measured a different location beyond the sharp
turn towards northwest in the stream). In general, there  is perhaps a slight gradient
in the visual colors towards blue as the aperture number increases, as seen in
Figure~\ref{magcolorfig}. For example, in $g-i$ the average of the median colors in
aperture 1 -- 15 is 1.3 and the average in apertures 16--23 is 0.9. While this
gradient is within the uncertainties of our colors, it could be consistent with
different progenitor star  populations that have been stripped at various times as
shown in the models of \citet{delgado08}. See also Fig.~\ref{visfig} in the current
paper where the right side shows with various colors stars that were stripped from the
simulated companion at various times. In the IRAC colors this trend is reversed and
the colors become slightly redder towards larger aperture numbers along the stream.
However, the largest gradients are seen in colors that include the Subaru $i$-band
that has the lowest surface brightness of any of the observed bands in the stream,
especially in apertures beyond aperture 23 that are therefore not plotted in
Fig~\ref{magcolorfig}. See Section~\ref{psf} below for a discussion about a
possible contribution to the color gradients from the extended PSFs. In general, the colors are roughly constant along the stream.

\subsection{Possible Effects of the Extended PSF}
\label{psf}

Formally, the very extended wings of the PSF could spread a small fraction of the 
light from NGC 5907 into a very faint and large-scale background gradient, dropping
away from the galaxy. Such a gradient, if present, could potentially affect the
brightness and colors derived for low surface brightness features such as the stellar
stream  studied here. To evaluate the effect of the extended PSF on our measurements, 
we have derived and adopted extended PSFs to construct models of the expected
background gradients.

To derive the radial profile of the extended halo light of the IRAC PSF, we started
by constructing the azimuthally averaged profile of \href{http://irsa.ipac.caltech.edu/data/SPITZER/docs/irac/calibrationfiles/psfprf/}{the warm IRAC extended PSF}.
It provides a radial profile extending to $\sim200\arcsec$. The profile at radii  from
$75\arcsec$ to $3000\arcsec$ was determined from serendipitous observations near $o$
Cet (Mira, program id 181), and observations around $\beta$ Gru (program id 1153) that
were specifically designed to elucidate the extended IRAC PSF. This very extended
profile was normalized  to match the radial profile of the above-mentioned extended
PSF at four points between 63\arcsec~ and 160\arcsec. The agreement in the region of
overlap is excellent (standard  deviation less than $\sim~4\%$), although the
normalization factor required means that  the faint ring in the profile at
$\sim1422\arcsec$ is about 1.7 times brighter than stated in the \href{http://irsa.ipac.caltech.edu/data/SPITZER/docs/irac/iracinstrumenthandbook/}{IRAC Instrument
Handbook} (see Section 7.3.4). By combining the two extended PSFs, the dynamic range of the IRAC
3.6 $\mu$m PSF was improved from about 2~$\times$~10$^{7}$ to about 10$^{10}$. The
full IRAC PSF to $3000\arcsec$ is shown in Figure \ref{fig:psf}.

An analysis of the bright stars in our Subaru images only yielded PSF profiles out to 
$\sim40\arcsec$, and did not show good consistency between cores
($0\arcsec-10\arcsec$), measured from unsaturated sources, and wings
($\sim5\arcsec-40\arcsec$), measured from much brighter, saturated  sources. A similar
analysis of the bright stars in the COSMOS field \citep{capak07} yielded a 
better consistency, but was still limited to radial extents of $\lesssim60\arcsec$.
Therefore, for the Subaru data, we adopted extended PSF profiles, generated from
observations made with other telescopes, from the review by \citet{sandin14}
and references therein. We tested two profiles, to check the sensitivity to the
adopted profile. The first tested profile was PSF$_{\rm{V,0m}}$ \citep{michard02},
augmented with PSF$_{\rm{K71}}$ \citep{king71} at radii $>150\arcsec$. We also
tested PSF$_{\rm{CV,1}}$ \citep{capa83} as an alternate PSF. The narrower core
of this PSF has a FWHM = $1\farcs 3$, which is a closer match to our measured Subaru
PSFs at FWHM $\approx$ $0\farcs 6$. More importantly for our models, the surface
brightness of this extended PSF is $\sim 3$ times brighter at 200\arcsec~ --
1000\arcsec~ than PSF$_{{\rm K71}}$. These PSFs are also shown in Figure
\ref{fig:psf}.

To model the background caused by the extended PSF in the IRAC mosaic, we worked
with  a shallower IRAC mosaic (from {\it Spitzer} program 3; see
Fig.~\ref{overlayfig}) that covered both the stream and the galaxy. We masked the
entire PID 3 mosaic except for pixels within 55\arcsec~ of the galaxy disk and within
30\arcsec~ -- 45\arcsec~ of the six brightest stars in the field. The stars were
included to check if the extended PSF from the bright stars might  have a significant
effect. We then convolved the masked mosaic with a 2-D symmetric kernel generated
from the very extended PSF profile in Figure~\ref{fig:psf}. The kernel is normalized
such that the total power is conserved in the convolved image. This generates the
model that is shown in grayscale and contours in Figure~\ref{fig:model1}. The
diffuse emission in the regions where the masked image is non-zero represents the
modeled extended PSF halo light from the galaxy and from the six brightest stars in
the field. In fact, the light from the extended PSF halo of the brightest stars is
found to have little effect on the overall extended PSF light predicted in the area
of the stream (see Fig.~\ref{fig:model1}). For each Subaru band, two similar models
were generated using the profiles in Figure~\ref{fig:psf}, and with the same mask as
applied for the IRAC model. The resulting Subaru model images are very similar to the
IRAC model, except that the light from the extended PSFs of the stars is less
prominent (i.e., underestimated) because the stars are saturated in the Subaru data.
The galaxy itself is only saturated in the central 4\arcsec~ $\times$ 10\arcsec~
region of the $i$-band image, which was filled by extrapolation from the $r$-band
image before calculating the light in the extended PSF. Slices through all the models
are shown in Figure~\ref{fig:slice}.

The models indicate that the expected levels of extended PSF halo light are about a
factor of ten (or more) below the background noise level of the images. Addition of
the  IRAC extended PSF model to the data has a clear impact on the scale height of
the galaxy and on the extended PSF emission surrounding the stars, but no apparent
effect at the distance where we measure the stellar stream ($>100''$ from the
galaxy). For the Subaru data, the PSF$_{\rm{CV,1}}$ model has a brighter and flatter
background at the $200\arcsec$ -- $600\arcsec$ radii relevant here.

We tested the effects of the extended PSFs on the measured surface brightnesses and
colors of the stellar stream by  adding them to the original images (we realize that
the extended PSF light component is already in the images, but here we are only
interested in the amount of change in the surface brightnesses and colors, so adding
instead of subtracting is justified just for this test). In the worst case, using the
less steeply falling PSF (PSF$_{\rm{CV,1}}$) for Subaru data, the surface
brightnesses changed by 0.21 mag (all the values listed here and in the following are
the median surface  brightnesses in magnitudes per pixel) at the very northwestern
end of the visible stream  (in apertures beyond 23 where the stream is very faint) in
the $i$-band. However, along most of the stream (apertures 3 -- 23) the change was
0.1 mag or less, and even less for the PSF$_{\rm{K71}}$ and the IRAC extended PSF.

\begin{figure*}
\centering
%\epsscale{0.85}
\includegraphics[angle=270,scale=0.6]{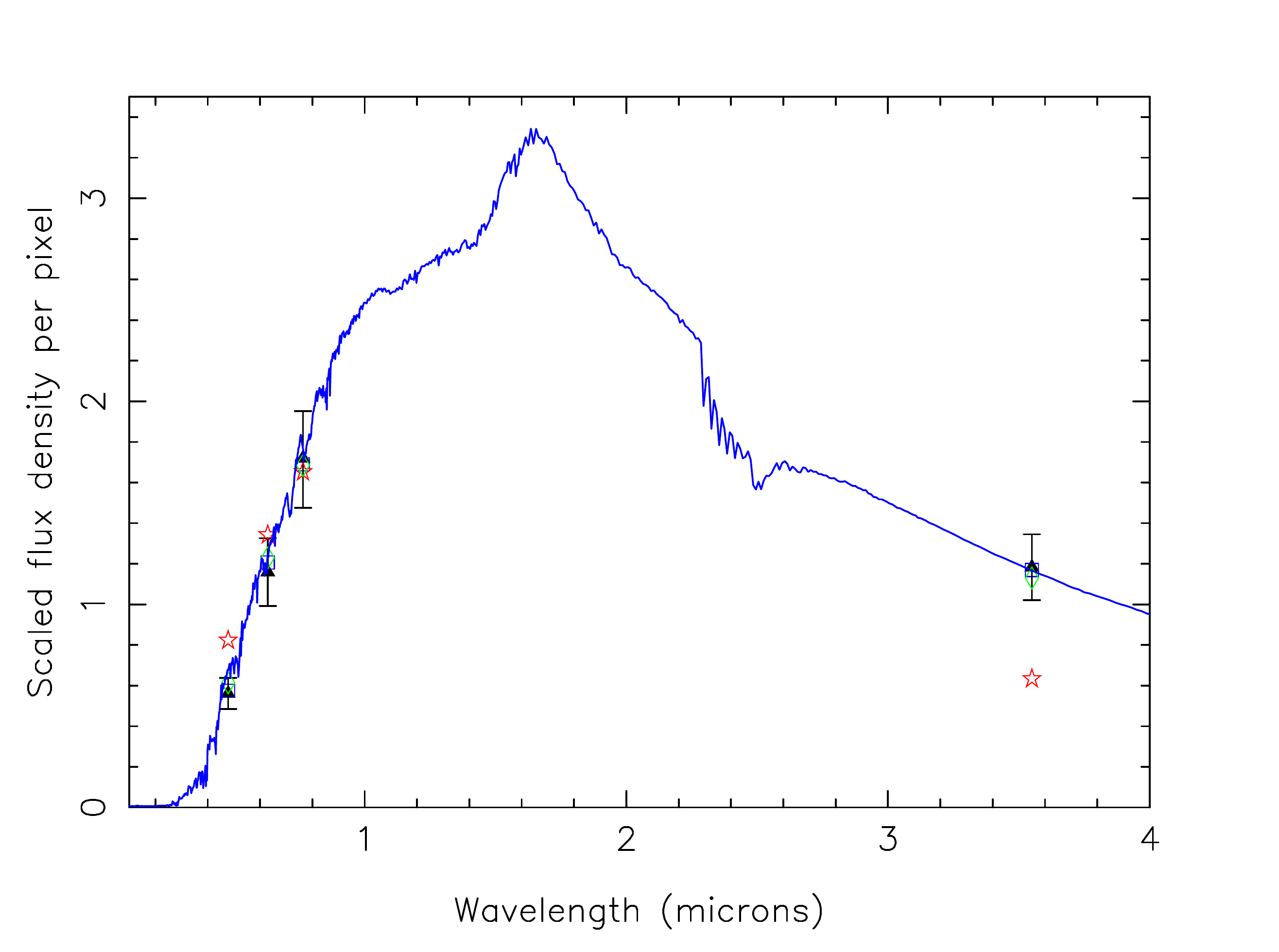}
\caption{Best fit of an FSPS/Padova model to the stellar stream SED around NGC 5907.
The observations are shown with black filled triangles. The error bars reflect the
uncertainty due to the background subtraction in the Subaru Suprime-Cam observations
and the uncertainty in the background determination in the IRAC 3.6 $\mu$m mosaic.
The best fitting FSPS/Padova model (age 14.96 Gyr and metallicity $-0.3$) is shown
with blue open squares and the corresponding model spectrum with a solid blue line. A
low-metallicity model (age 9.44 Gyr and metallicity $-1.98$, to demonstrate that low metallicity models are ruled out) is shown with open red stars, and a model with an age of 10 Gyr and metallicity of $-0.3$ with open green diamonds. The y-axis is in arbitrary surface brightness units.\label{sedfig}} 
\end{figure*}

\begin{figure}
\centering
%\epsscale{0.85}
\includegraphics[angle=270,scale=0.35]{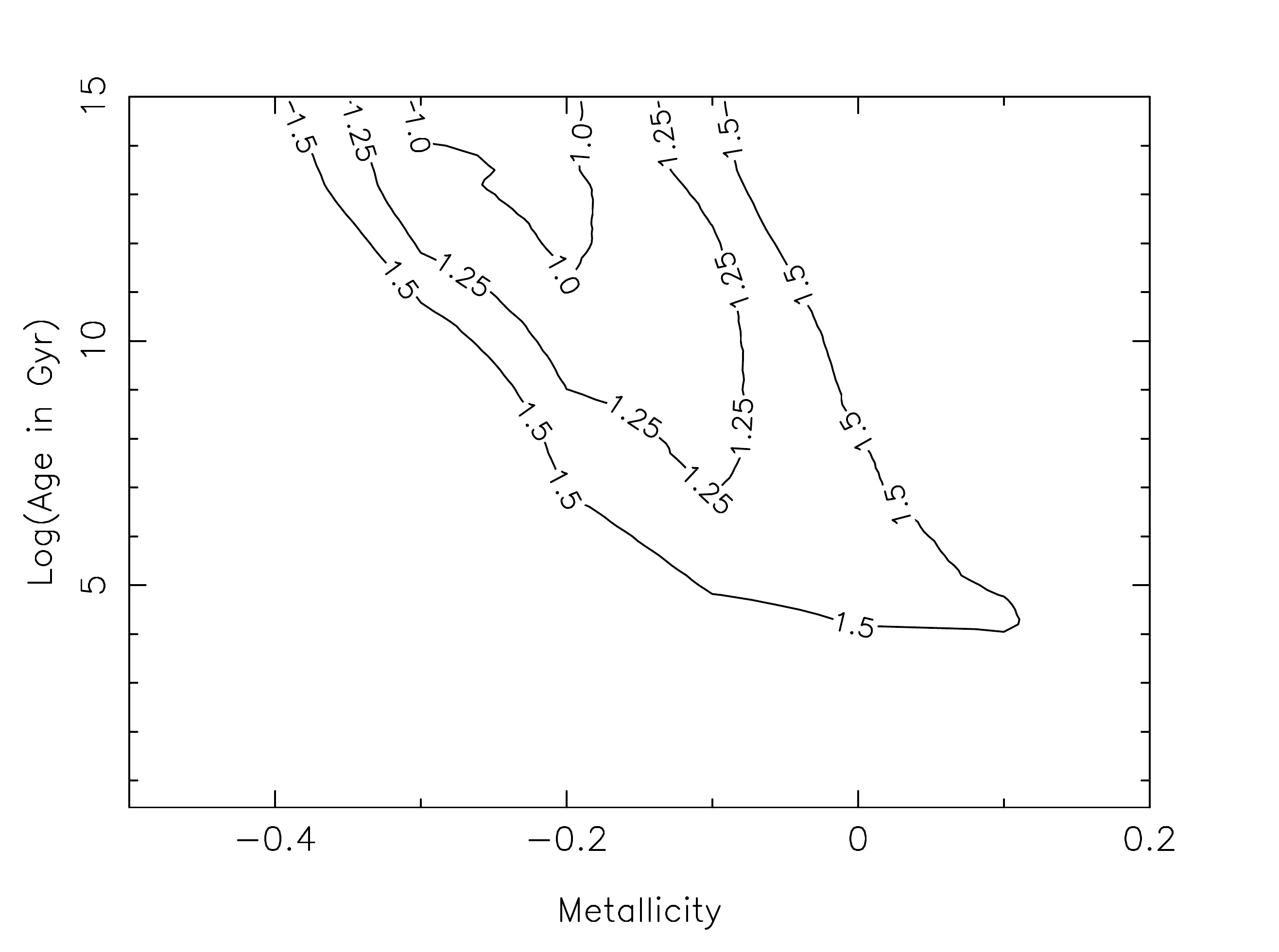}
\caption{$\chi^{2}$ values of the fits of the FSPS/Padova models to the data, in age -- metallicity space. The contours correspond to about 87\% ($\chi^{2}$ = 1.25) and  83\% ($\chi^{2}$ = 1.5) confidence limits for four degrees of freedom.\label{chifig}} 
\end{figure}

We also checked the changes in broad-band colors along the stream after adding the
extended PSF contribution. At worst, the visible -- IRAC colors changed by about 0.05
mag (became bluer) along most of the stream (apertures 3 -- 23) and the colors
changed even less for the visible -- visible bands, with no gradient in these changes
along the stream. Making the IRAC surface brightness dimmer by 0.05 magnitudes with
respect to the Subaru band surface brightnesses does not change our best fit results
for metallicity and age below

\section{Spectral Energy Distribution Fitting}
\label{sed}

We calculated the median surface brightnesses in the apertures defined in
Section~\ref{magcolor}. Because the stream is getting very faint in at least the
$i$-band after its easternmost point (and towards the northwest from there), we
calculated the average of the median aperture flux densities in each band, 
taking into account only apertures 1--23. 

We constructed models of single-burst stellar populations with varying ages and
metallicities using the flexible stellar population synthesis (FSPS) code of
\citet*{conroy09} and \citet{conroy10}. FSPS makes use of both the Padova
\citep{girardi00,marigo07,marigo08} and a ``bag of stellar tracks and isochrones''
\citep[BaSTI;][]{pietri04,cordier07} stellar evolution calculations in the
versions available as of 2009 November. We also used FSPS's Python bindings
\citep*{foreman14}. The range in metallicity (log~[Z/Z$_{\sun}$]) in the FSPS/Padova
models is from $-1.98$ to $+0.2$ and in the FSPS/BaSTI models from $-1.80$ to $+0.32$.
The age range that we used was from 0.47 Gyr to 14.96 Gyr for the FSPS/Padova models
and from 1.86 Gyr to 14.79 Gyr for the FSPS/BaSTI models. By scaling the models we
achieved the best least-squares fit to the four average flux densities per pixel for
an FSPS/Padova model with a stellar population age of 14.96 Gyr and a metallicity of
$-0.3$, with a 95\% confidence limit on the metallicity of $-0.3$$\pm$0.2. The best
fit is shown in Figure~\ref{sedfig}, together with an FSPS/Padova model having an age
of 9.44 Gyr and a low metallicity log~[Z/Z$_{\odot}$] of $-1.98$ (showing that old
age/low metallicity stellar populations are ruled out). The corresponding $\chi$$^{2}$
values of the best fit are shown in Figure~\ref{chifig}. The lowest $\chi$$^{2}$
value for the FSPS/BaSTI models gives a stellar population age around 12 Gyr and a
metallicity near $-0.3$. It should be noted that including the IRAC 3.6 $\mu$m band is
crucial for the metallicity determination. Averaging the first 12 apertures, the best
fitting model has a metallicity of $-0.5$ and again an age of 14.96 Gyr, and averaging
apertures 13--23 gives a best metallicity fit of $+0.2$ and age of 3.6 Gyr. These
trends in age and metallicity may cause the colors to remain roughly constant. It is
possible that despite vigorous masking, there is leftover light from the foreground
stars that causes the young metal-rich stellar population to show up in apertures
13--23 when averaged together, as can be seen in Figure~\ref{boxfig} (where there
were bright foreground stars in the apertures before masking). Excluding the possibly
problematic $i$-band (see Section~\ref{estmagcolor}) from the SED fitting does not
change the best fitting stellar population parameters appreciably (best-fitting  age
11.9 Gyr, best-fitting metallicity $-0.2$).

\begin{deluxetable*}{lccc}
\tabletypesize{\scriptsize}
\tablecaption{ESTIMATED PROPERTIES OF THE IMAGED STREAM SECTION\label{table1}}
\tablewidth{0pt}
\tablehead{
\colhead{Property} & \colhead{Bandpass} & \colhead{Value} & \colhead{Unit} \\
}
\startdata
Integrated absolute magnitude  & $g$ & $-14.1$ & mags \\
Integrated absolute magnitude  & $r$ & $-14.8$ & mags \\
Integrated absolute magnitude  & $i$ & $-15.2$ & mags \\
Integrated absolute magnitude  & $3.6$ & $-14.6$ & mags \\
Luminosity & $g$ & 4.9~$\times$~10$^{7}$ & $L$$_{\sun}$ \\
Luminosity & $r$ & 6.0~$\times$~10$^{7}$ & $L$$_{\sun}$ \\
Luminosity & $i$ & 7.8~$\times$~10$^{7}$ & $L$$_{\sun}$ \\
Luminosity & $3.6$ & 1.7~$\times$~10$^{8}$ & $L$$_{\sun}$ \\
Mean $g-r$ & $g,r$ & 0.7 & mags \\
Mean $g-i$ & $g,i$ & 1.0 & mags \\
Mean $r-i$ & $r,i$ & 0.3 & mags \\
Mean $g-3.6$ & $g,3.6$ & 0.9 & mags \\
Mean $r-3.6$ & $r,3.6$ & 0.1 & mags \\
Mean $i-3.6$ & $i,3.6$ & $-0.1$ & mags \\
Stellar mass & $r$ & 2.1~$\times$~10$^{8}$ & $M_{\sun}$ \\
Mean projected distance to the center of NGC 5907 & $r$ & 9.4/46 & arcmin/kpc \\
Detected length of stream & $r$ & 11.7/57.8 & arcmin/kpc\\
\enddata

\tablecomments{All the estimates are derived from the apertures shown in
Fig.~\ref{boxfig}. All the magnitudes and colors are in the  AB system. Values are
measured for apertures 1--33. The absolute magnitudes are calculated by integrating
the flux densities in all the measured apertures and summing them up. The luminosities
were calculated using absolute magnitude values of the Sun in the AB system as
tabulated in FSPS (5.12, 4.64, 4.53, and 5.99 for $g$, $r$, $i$, and 3.6 $\mu$m,
respectively). The mass was calculated from the $r$-band luminosity and the $M$/$L$
ratio (3.46, assuming a \citeauthor{kroupa01} \citeyear{kroupa01} type initial mass
function, IMF) of the best fitting model in Section~\ref{sed}. The mean projected
distance was calculated as the mean of the distances of the centers of each aperture
box 1--33. The detected length is from the center of aperture 1 through the centers of
apertures 2--33 and the physical value in kiloparsecs (kpc) is assuming a distance of
17 Mpc.}

\end{deluxetable*}

\begin{figure*}
\centering
%\epsscale{0.85}
\includegraphics[angle=0,scale=0.75]{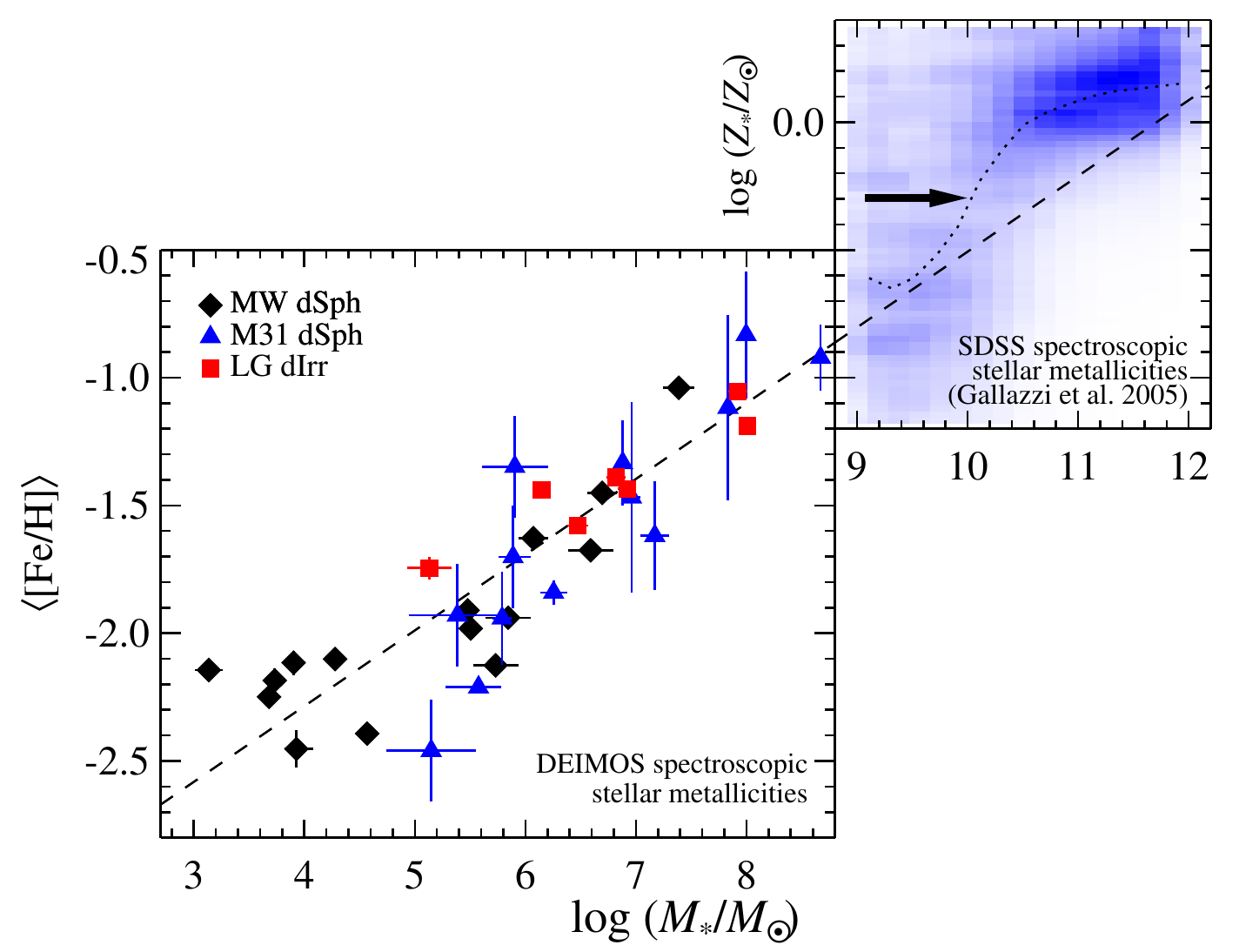}
\caption{Stellar mass -- stellar metallicity relation, reproduced from Figure 9 of
``The Universal Stellar Mass--Stellar Metallicity Relation for Dwarf Galaxies'' by E.
N. Kirby,  J. G. Cohen, P. Guhathakurta, L. Cheng, J. S. Bullock, \& A. Gallazzi, ApJ,
vol 779, Issue 2 (2013), pp. 102--122. The lower left shows the relation for Local
Group dwarf galaxies and the upper right for more  massive galaxies in SDSS from
\citet{gallazzi05}. The dotted line in the diagram for the more massive galaxies is
the median of the stellar metallicity distribution as a function of stellar mass. The
black arrow was added to point to the metallicity value measured in this work for
NGC~5907. See \citet{kirby13} for more information.\label{mass-metalfig}} 
\end{figure*}

The age of the best fitting model may seem problematic because the current best
estimate of the age of the Universe is about 13.8 Gyr \citep{planck15}. Basically
the fit is telling us that the oldest possible stellar population provides the best
fit. A fit with an age of 10 Gyr is shown with diamonds in Figure~\ref{sedfig} and
is only slightly worse than the fit with the unphysically large age. The stellar
populations in the Sagittarius stream \citep{chou07} and M31's giant stream also have
old ages \citep[older than about 7 Gyr;][]{brown06,tanaka10,deboer15}.

It should be kept in mind that there are well-known challenges with modeling
stellar populations in the infrared. For example, the flux contributions from
thermally-pulsing asymptotic giant branch stars at 3.6~$\mu$m are uncertain
\citep[cf.][]{villaume15}, and if high, they could shift our solution to a
somewhat younger age and lower metallicity. In addition, there is a general tendency
for stellar population synthesis models, including FSPS, to predict overly blue
visible-light to near-infrared colors \citep[e.g.][]{conroy10}, which means that
our metallicity inference could be too high. On the other hand, our  inference was
driven not only by the near-infrared, but also by the relatively red  optical colors of
the stream.

\section{Discussion}
\label{discus}

The stellar stream in NGC 5907 has been previously modeled with a satellite galaxy
accretion event \citep[][total mass ratio 1:4000]{delgado08} and with a gas-rich major
galaxy merger \citep[][total mass ratio 1:3 -- 1:5]{wang12}. Neither N-body model
achieved a quantitative fit to the stream, but both were able to qualitatively create
features similar to those seen in the observations. 

The main objection to the satellite galaxy accretion model is that there is no 
obvious remnant of the nuclear region of the disrupted satellite \citep{wang12}.  In
NGC 5387 \citet{beaton14} found signs of a very blue star-forming cluster right at
the intersection of a stellar stream and the host galaxy disk. They speculated that
this could be the starbursting nuclear remnant of the mostly disrupted satellite
galaxy. However, no such star-forming region has been found near the  intersections
of the disk of NGC~5907 and the stellar stream, and we do not see any such blue star
forming region along the brightest section of the stellar stream that we have imaged.
Neither have we detected any signs of active star formation in the stellar stream.

The main objection to the major galaxy merger model is that NGC 5907 appears to have
a thin disk with no obvious bulge \citep[cf.,][]{lia16}, and does not look like a
severely perturbed major merger remnant (see major merger simulations by, e.g.,
\citeauthor{barnes92} \citeyear{barnes92}). In the gas-rich major merger scenario,
the thin disk of NGC 5907 is rebuilt around a small bulge. A similar claim for the
creation of the faint stellar stream features around NGC~4013 by a major merger has
recently been made by \citet{wang15}. Another objection to the major merger model
could be the lack of any detected neutral  atomic hydrogen (\ion{H}{1}) gas in the
stellar stream \citep{shang98}, if it is interpreted as being a long remnant tidal
tail.

Our results suggest that the stream in NGC 5907 is relatively metal-rich: the best
fitting FSPS model has a metallicity log~[Z/Z$_{\sun}$] ($\approx$~[Fe/H], assuming
the fractional percentage of hydrogen is the same as in the Sun and that the iron
abundance and the total metal abundances are roughly the same) = $-0.3$. This should
be compared to the reported metallicity in the Sagittarius stream around the Milky
Way, from a peak [Fe/H] of $-0.3$ in its core to a median [Fe/H] from $-0.7$ to
$-1.1$ in its leading arm \citep{chou07}. The Sagittarius stream is known to be an
interacting satellite galaxy, and its core has been identified. While the stellar
mass of the Sagittarius dwarf (the progenitor of the Sagittarius stream) is currently
about $2.5 \times 10^{8} M_{\sun}$, the N-body modeling of \citet{law10} suggests an
original mass of $6.4 \times 10^{8} M_{\sun}$. Meanwhile, in the giant southern
stellar stream of M31, the [Fe/H] has a strong peak at $-0.3$, with a mean of $-0.55$
and a median of $-0.45$ \citep{kalirai06,guhatha06,tanaka10}. The disrupted precursor
galaxy for the giant southern stellar stream in M31 could have been a satellite of
stellar mass 4~$\times$~10$^{9}$~M$_{\sun}$ \citep{dekel03,tanaka10}. A comparison to
the $g-i$ color of about 0.6 -- 0.8 for the umbrella in NGC~4651 \citep{foster14}
that is supposedly a result from a merger of a metal-poor satellite suggests that the
merging galaxy in the NGC 5907 system was more metal-rich. 

Using the metallicity of the NGC\,5907 stellar stream and the metallicity vs. stellar
mass relationship in \cite{dekel03} and \citet[][reproduced in
Fig.~\ref{mass-metalfig} of the current paper]{kirby13}, the disrupted companion of
NGC\,5907 would have had a stellar mass of about 1~$\times$~10$^{10}$~M$_{\sun}$, with
a large uncertainty that extends down to $\sim$~1~$\times$~10$^{9}$~M$_{\sun}$. Note
that the mass -- metallicity relationship evolves when going to a higher redshift
sample. Similarly, equation (9) in \citet{kirby11} predicts  $L_{\rm tot} = 4.1\times
10^{10} L_{\sun}$ for the disrupted companion of NGC 5907,  using our derived [Fe/H]
value of $-0.3$. This would make this system qualify as a major merger, because the
dynamical disk mass in NGC 5907 has been estimated to be about
1.4~$\times$~10$^{11}$~M$_{\sun}$ \citep[][adjusted for the adopted distance of 17 Mpc
in our study]{casertano83}, of which gas may constitute about 10\% \citep{dumke97}.
\citet{just06} modeled the disk of NGC 5907  with a stellar mass of
2~$\times$~10$^{10}$~M$_{\sun}$, making the merger an almost  equal disk mass merger.
However, using the conversion from IRAC 3.6 and 4.5 micron flux densities by
\citet*{eskew12}, the stellar mass of  NGC 5907 is about
7.8~$\times$~10$^{10}$~M$_{\sun}$. We adopt this as the best  estimate to compare with
our (partly) IRAC-based metallicity measurement. 

We have estimated the mass in the IRAC-imaged part of the stellar stream
(Table~\ref{table1}). The value we obtain, 2.1~$\times$~10$^{8}$ $M_{\sun}$, is
consistent with the value estimated by \citet{delgado08}, 3.5~$\times$~10$^{8}$
$M_{\sun}$, because \citet{delgado08} used the luminosity of the whole stream that is
seen in Fig.~\ref{visfig} to estimate the mass. This mass would imply a satellite
accretion event that may also be consistent with our color findings if one allows
for systematic issues in the stellar population modeling (see Section~\ref{sed}). On
the other hand, the estimated stream mass is only a lower limit to the progenitor
mass and thus it cannot accurately constrain the merger scenario, because in a
minor merger most of the merged companion mass has presumably become part of the
bulge and disk of the primary galaxy, NGC~5907.

\section{Conclusions}
\label{concl}

We have studied the brightest part of the stellar stream around NGC~5907 with both
visible-light ($g$, $i$, and $r$) and IRAC near-infrared 3.6 $\mu$m images. While
the stream is very faint, we have managed to generate color index profiles and an SED
for the brightest part of the stream. Our results show a slight color change in some
of the bands along the stream, most notably in $g-i$ and $r-i$ in which the color 
becomes bluer after the turning point in the northeast corner of the brightest  parts
of the stream. However, there may be a remaining background subtraction problem in
the $i$-band and our results are essentially consistent with no color gradients along
the stream.

We have fitted FSPS stellar population synthesis models with a single-age stellar 
population. The best fit is achieved for an old and relatively metal-rich stellar
population in the stream. Such a stellar population appears to be more consistent
with a rather massive minor merger event than with the stripped stars from a dwarf 
galaxy accretion event. 

Future work should include better visible light photometry of the stream, and
improving the stellar population synthesis model predictions in the near-infrared.
Future kinematical studies of globular clusters associated with this stream could 
shed light on the implied minor merger scenario (Alabi et al., in preparation). 

\acknowledgments

We thank the referee, Rodrigo Ibata, for critical comments that greatly improved
the paper. We are indebted to David Shupe for help with fitting the stellar
population synthesis models. We thank Charlie Conroy, Ben Johnson and Alexa
Villaume for helpful discussions regarding stellar population synthesis codes. We
thank Ben Johnson for his help with the Python interface to FSPS. We thank Sean Carey
for useful discussions  on error estimation. And we thank Evan Kirby and Anna
Gallazzi for permission to reproduce a figure from their paper. DMD was supported by
Sonderforschungsbereich  SFB 881 "The Milky Way System" (subproject A2) of the German
Research Foundation (DFG). This research was supported in part by the National Science
Foundation under grant Nos. AST-1211995 and AST-1518294. This work is based in part on
observations made with the Spitzer Space Telescope, which is operated by the Jet
Propulsion Laboratory, California Institute of Technology under a contract with NASA.
Support for this work was provided by NASA through an award issued by JPL/Caltech.
Based in part on data collected at Subaru Telescope, which is operated by the National
Astronomical Observatory of Japan. This research has made use of the NASA/IPAC
Extragalactic Database (NED) which is operated by the Jet Propulsion Laboratory,
California Institute of Technology, under contract with the National Aeronautics and
Space Administration. This research has made use of the NASA/ IPAC Infrared Science
Archive, which is operated by the Jet Propulsion Laboratory, California Institute of
Technology, under contract with the National Aeronautics and Space Administration.

\vspace{5mm}

{\it Facilities:} \facility{Spitzer,Subaru (Suprime-Cam)}

\software{MOPEX, SExtractor, Adobe Photoshop}

\end{document}